\documentstyle[aps,prl,floats,epsf,twocolumn]{revtex}
\newcommand{\be}{{\bf e}}
\newcommand{\br}{{\bf r}}
\newcommand{\bk}{{\bf k}}
\newcommand{\bq}{{\bf q}}

\begin{document}
\draft
\twocolumn[\hsize\textwidth\columnwidth\hsize\csname@twocolumnfalse%
\endcsname

\draft
\preprint{}

\title{Vortex glass transition in a random pinning model}

\author{Anders Vestergren$^{(a)}$, Jack Lidmar$^{(b)}$, and Mats
Wallin$^{(a)}$}

\address{(a) Condensed Matter Theory, Royal Institute of Technology, SCFAB,
SE-106 91 Stockholm, Sweden\\ (b) Department of Physics, Stockholm
University, SCFAB, SE-106 91 Stockholm, Sweden}

\maketitle
\begin{abstract}
We study the vortex glass transition in disordered high temperature
superconductors using Monte Carlo simulations.  We use a random
pinning model with strong point-correlated quenched disorder, a net
applied magnetic field, longrange vortex interactions, and periodic
boundary conditions. From a finite size scaling study of the helicity
modulus, the RMS current, and the resistivity, we obtain critical
exponents at the phase transition.  The new exponents differ
substantially from those of the gauge glass model, but are close to
those of the pure three-dimensional XY model.
\end{abstract}
]

The magnetic field-temperature phase diagram for vortices in
disordered high temperature superconductors has been the focus of
considerable interest in the past few years, centered around e.g. the
suggestion of M.\ P.\ A.\ Fisher of a possible vortex glass phase with
vanishing linear resistance \cite{Fisher}.  Numerical simulation is
often an important tool in investigations of phase transitions in
vortex systems.  For example, recent simulations have provided
valuable information about the phase diagram for the case of weak
disorder or weak magnetic fields \cite{Olsson}.  Here evidence for a
first order transition separating a Bragg glass phase
\cite{Bragg-glass}, i.e.\ a dislocation-free solid with algebraically
decaying translational order, from a vortex liquid phase was obtained.
However, in the opposite limit of strong disorder or strong magnetic
fields, the existence of a vortex glass phase, i.e.\ a vortex solid
which is topologically disordered in terms of frozen in dislocations,
and the critical properties of the vortex glass transition, have not
been settled.  In this paper we study these issues by Monte Carlo
simulations and finite size scaling.

Simulations of disordered superconductors have often used a so-called
gauge glass model \cite{Olson-Young}, where the disorder is modeled as
a random vector potential added to the phase difference of the
superconducting order parameter.  This disorder corresponds to
spatially random magnetic flux present in the system.  Interestingly,
recent simulation results for the gauge glass \cite{Olson-Young} gives
similar values for the critical exponents as obtained in certain
recent experiments where a vortex glass has been reported
\cite{vortexglass-experiment}.  However, the gauge glass model has two
features that are not particularly realistic, namely, it does not
assume any pinning directly affecting the vortex core energy, and,
secondly, it is completely isotropic and does not contain any net
magnetic field.  It is at present unknown if these details are
important for the universality class of the glass transition, i.e.\ if
they modify the critical exponents.  In particular, there is a
possibility of anisotropic scaling produced by the net field, such
that the correlation length diverges with different exponents along
and perpendicular to the field.  Some of these and other issues have
been addressed in the literature.

An XY model with a random coupling constant and a net magnetic field
has been simulated using open boundary conditions \cite{Kawamura}.
However, periodic boundary conditions are preferable when bulk
properties are studied, to avoid any influence of the sample surface.
Also the effect of screening of the vortex interaction has been
considered.  Gauge field fluctuations lead to screening of the vortex
interaction.  This screening is usually rather weak in the high
temperature superconductors.  In models for the strong screening
limit, simulations give evidence for the absence of a vortex glass
phase at finite temperatures \cite{screening}.  A vortex glass
scenario without any thermodynamic phase transition has also been
suggested \cite{Zimanyi}.

In this paper we consider a random pinning model that contains all the
pieces necessary to describe the static universal critical properties
of the vortex glass critical point: long-range vortex interactions,
strong vortex pinning, a net applied magnetic field, and periodic
boundary conditions.  The vortex-vortex interaction is a full 3D
longrange interaction without screening, that becomes applicable when
the bare screening length is much longer than the vortex spacing,
i.e.\ in the strong field limit.  The vortex pinning corresponds to
uncorrelated quenched point disorder, implemented as a position
dependent core energy.  This is equivalent to random-$T_c$ disorder.
The vortex-vortex interaction and the disorder are isotropic in the
model, but the applied net magnetic field breaks the spatial symmetry.
We include the possibility of anisotropic scaling by allowing for
different correlation length exponents in the directions parallel and
perpendicular to the magnetic field.  The dynamic universality class
assumed here is that of relaxation dynamics of the vortex lines, and
we stress that there are more dynamic universality classes possible
\cite{dynamics}.

The random pinning model is defined by the Hamiltonian
\begin{equation}
H=\frac{1}{2} \sum_{\br, \br'} V(\br-\br') \bq(\br) \cdot \bq(\br') + \\ 
\frac{1}{2} \sum_{\br, \mu} D_\mu(\br) q_\mu(\br)^2
\end{equation}
The model is defined on a simple cubic lattice with $\Omega=L\times
L\times L_z$ sites, using periodic boundary conditions in all three
directions.  The vortex line variables are specified by an integer
vector field $\bq(\br)$, whose $\mu=x,y,z$ component is the vorticity
on the link from $\br$ to $\br+\be_\mu$.  The partition function is
$Z={\rm Tr} \exp(-H/T)$, where $T$ is the temperature, and ${\rm Tr}$
denotes the sum over all possible integers $q_\mu$, subject to the
constraint $\nabla \cdot \bq=0$ on all sites, i.e., the vortex lines
have no open ends.  An applied net magnetic field is included as a
fixed number of vortex lines penetrating the system in the
$z$-direction.  Three different fillings of the applied magnetic field
are considered here, i.e., number of vortex lines per link in the
$z$-direction: $f=1/2, 1/4, 1/\sqrt{10}$.  For the irrational filling
we used the integer number of vortex lines closest to $fL^2$.  The
figures below are for $f=1/2$.  The long range vortex-vortex
interaction is given by
\begin{equation}
V(\br)=\frac{K}{\Omega}\sum_{\bk} \frac{e^{i \bk \cdot
\br}}{\sum_\mu (2-2\cos k_\mu )}
\end{equation}
where $K=4\pi^2J$ (we set $J=1$).  On each link in the system is a
short-range point-correlated random pinning energy with a uniform
distribution in the interval $0 \le D_\mu (\br) \le K$.
Hence the lattice constant of the discretization lattice in our model
corresponds physically to a characteristic length scale for variations
in the disorder energy landscape.

The Monte Carlo (MC) trial moves are attempts to insert vortex loops
with random orientation on randomly selected plaquettes of the
lattice.  A MC sweep consists of one attempt on average to insert a
loop on every plaquette.  The attempts are accepted with probability
$1/(1+\exp \Delta E/T)$, where $\Delta E$ is the energy change for
inserting the loop.  For half and quarter fillings the initial vortex
configurations consist of straight lines along the $z$-direction,
arranged in a regular lattice in the $xy$-plane.  For filling
$f=1/\sqrt{10}$ the initial configuration has straight lines placed at
random.  For equilibration about $10^5$ MC sweeps are used, followed
by equally many sweeps for collecting data.  The results were averaged
over up to 2000 samples of the disorder.  Thermal averages are denoted
by $\langle \cdots \rangle$ and disorder averages by $[ \cdots ]$.  To
avoid systematic errors in the calculation of squares of expectation
values, two replicas of the system with the same disorder are used.

Superconducting coherence in the vortex line system can be detected by
calculating the helicity modulus.  One way of doing this
\cite{helicity} is to add a term $H_Q=\frac{K}{2 \Omega} {\bf Q^2}$ to
the Hamiltonian, where $Q_\mu$ is the total projected area of vortex
loops added during the simulation.  The helicity modulus in the
direction $\mu$ is then given by
\begin{equation}
\label{upsilon}
Y_\mu=1-\frac{K}{\Omega T} [ \langle Q_\mu^2 \rangle - \langle
Q_\mu \rangle^2 ]
\end{equation}
and the RMS current density is defined as
$J_\mu=\frac{K}{\Omega} [ \langle Q_\mu \rangle^2 ]^{1/2}$.
The linear resistivity, $\rho$, is obtained from the Kubo formula for
the resistance:
\cite{Kubo}
\begin{equation}
\label{rho} 
R_\mu=\frac{1}{2T} \sum_{t=-t_0}^{t_0} [ \langle
V_\mu(t) V_\mu(0) \rangle ]
\end{equation}
where $t$ is MC time, and the voltage is $V_\mu \propto \Delta Q_\mu$
is the net change in the projected vortex loop area during a sweep.
In the calculation of $\rho$ the $H_Q$ term is not included in $H$.
The summation time $t_0$ is chosen large enough that the resistivity
is independent of $t_0$, but much shorter than the length of the
simulation.

We use a generalization of the Fisher-Fisher-Huse \cite{Fisher}
scaling ansatz to analyze our MC data.  At the glass transition
temperature $T_c$ the correlation length in the $xy$-planes, $\xi$,
and in the $z$-direction, $\xi_z$, and the correlation time, $\tau$,
are assumed to diverge as $\xi \sim |T-T_c|^{-\nu}, \xi_z \sim
\xi^\zeta$, and $\tau \sim \xi^z$.  The anisotropic finite size
scaling ansatz \cite{Boseglass} for the helicity modulus is
\begin{eqnarray}
\label{fss_Yx} Y_x&=&L^{3-d-\zeta}
f_x(L^{1/\nu}(T-T_c),L_z/L^\zeta)
\\ \label{fss_Yz}  Y_z&=&L^{1-d+\zeta} f_z(L^{1/\nu}(T-T_c),L_z/L^\zeta),
\end{eqnarray}
where $d=3$ is the spatial dimensionality and $f_{\mu}$ are scaling
functions (scaling functions will from now on be suppressed in the
equations).  The current density scales as $J_x \sim L^{2-d-\zeta},
J_z \sim L^{1-d}$.  The linear resistivity $\rho=E/J$, where $E$ is
the electric field, obeys
\begin{equation}
\label{fss_rho}
\rho_x \sim L^{d-3+\zeta-z},
\rho_z \sim L^{d-1-\zeta-z},
\end{equation}

We did a number of tests of equilibration of our simulations.  We
followed the standard procedure of calculating the ``Hamming
distance'' between two replicas with identical disorders
\cite{dirtybosons}, for some selected system sizes $L,L_z$.  We also
increased the number of sweeps for equilibration with a factor of 10
for a few selected parameter values, and obtained no deviations from
the data in the figures below.  The linear resistivity (shown below in
Fig.\ \ref{rhofig}) gives an estimate of the relaxation time $t_0$,
which is the time where the curves saturate.  The values for $t_0$
gives a rough estimate of the required equilibration time.  The
equilibration times used in our simulations are $\approx 8 t_0$.

\begin{figure}[htb]
\centerline{\epsfxsize 6cm\epsfbox{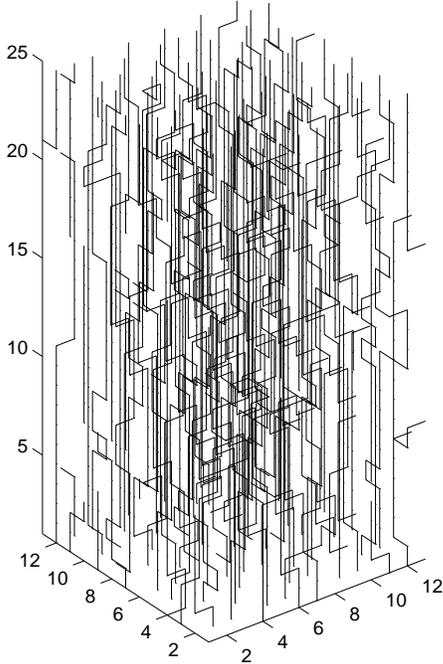}}
\caption{Snapshot of a vortex line configuration from our MC
simulation for $f=1/2$ at $T=0.5$ ($\ll T_c$) which is deep in the
vortex glass phase.}
\label{snapshot}
\end{figure}

The first step is to verify that the model has a vortex glass phase,
instead of a Bragg glass phase \cite{Bragg-glass} that is expected for
low fillings and weak disorder.  Figure \ref{snapshot} shows a typical
snapshot of a sample configuration for $T=0.5 (\ll T_c)$, i.e.\ deep
in the glass phase.  This calculation was done using an exchange MC
algorithm \cite{exchange} with 9 uniformly spaced temperatures in
$[0.5 , 4.5]$.  We also computed the structure function $S({\bf q})$,
and obtain no essential difference between $S$ in the vortex liquid
phase and the glass phase, with no indication of a Bragg glass phase
\cite{Bragg-glass}.  This demonstrates that the low temperature phase
of our model, for the filling and disorder strength considered, is
indeed a glass where the vortex lines are frozen in random positions.

\begin{figure}[htb]
\centerline{\epsfxsize 7cm\epsfbox{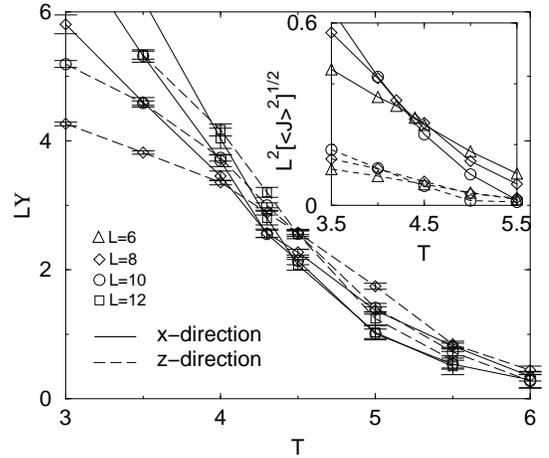}}
\caption{MC data for the helicity modulus vs.\ temperature for
different system sizes $L$.  Inset: RMS current density vs.\
temperature.}
\label{helicity}
\end{figure}

To locate the critical temperature of the vortex glass to liquid
transition, and determine the critical exponents, MC data for the
helicity modulus in the $x$ and $z$ directions is analyzed by the
finite size scaling form in Eqs.\ (\ref{fss_Yx}),(\ref{fss_Yz}).  MC
data for different $T, L, L_z$ for $f=1/2$ is plotted in Fig.\
\ref{helicity}.  To determine the anisotropy exponent $\zeta$, a
sequence of system sizes $L_z$ are examined for each $L$.  In the
relation $L_z=cL^\zeta$, both $c$ and $\zeta$ are varied until data
curves for $Y_x$ and $Y_z$ for different system sizes $L=6,8,10,12$
become approximately independent of $L$ at a common temperature, which
is our estimate for $T_c$.  Fig.\ \ref{helicity} shows the best
crossing obtained in this procedure, which gives $T_c = 4.5 \pm 0.1,
\zeta= 1 \pm 0.1$.  The same results are obtained from the RMS
currents (inset).  Equally good fits are obtained for $c$ in the
interval $1.5 < c < 2$.  The errors are estimated as the interval
outside which considerably poorer scaling is obtained.

\begin{figure}[htb]
\centerline{\epsfxsize 7cm\epsfbox{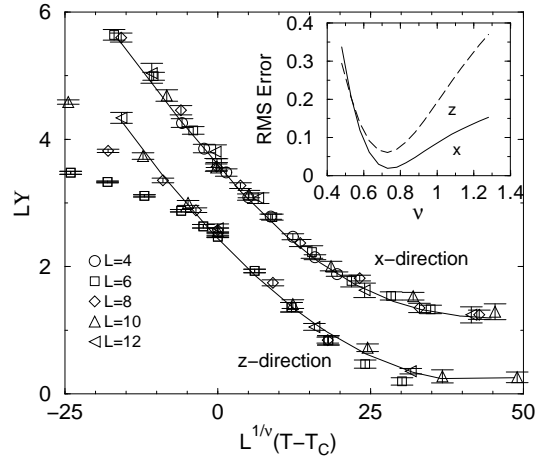}}
\caption{
Finite size scaling data collapse of MC data for the helicity moduli
in the $x$ and $z$-directions, obtained by fitting both $T_c$ and
$\nu$.  The data in the $x$-direction has been shifted to $LY_x+1$ for
clarity.  Solid curves are guides to the eye.  Inset: RMS fit error
vs.\ $\nu$ at $T_c=4.5$.}
\label{nu}
\end{figure}

To determine the correlation length exponent $\nu$ we use fits to
Eqs.\ (\ref{fss_Yx}),(\ref{fss_Yz}) to obtain a data collapse for
system sizes $L=6-12$ over an entire temperature interval around
$T_c$.  As a measure of the quality of the collapse we define the RMS
fit error $\Delta=\left[ \sum_{L,T,\mu} (LY_\mu(L,T)-f_\mu(x))^2
\right]^{1/2}$, where $x=L^{1/\nu}(T-T_c)$, and $f_\mu$ is estimated
by e.g.\ a cubic polynomial fit to the MC data.  We did several
different types of fits, all giving similar results.  Figure \ref{nu}
shows the best data collapse in a two parameter fit where both $T_c$
and $\nu$ are varied independently in the $x$ and $z$-directions.  The
best fit is obtained for $T_c=4.5 \pm 0.1, \nu=0.7 \pm 0.1$.  A good
data collapse is obtained, except for the smallest system size, where
deviations are obtained below $T_c$.  However for larger sizes scaling
gets better, suggesting that the deviation for small system size is a
finite size effect.  The inset shows the RMS fit error as a function
of $\nu$ for fixed $T_c=4.5$.  Nearly identical results are obtained
by instead analyzing data for the RMS currents.  The same result for
$\nu$ is also obtained from an analysis of the derivative of the
helicity moduli wrt $T$.  For the lower fillings, $f=1/4,
1/\sqrt{10}$, we find similar exponents but with larger error bars.

So far we have presented results for static quantities, and we now
consider dynamics.  The dynamic critical exponent is obtained from the
linear resistivity in Eq.\ (\ref{rho}), for $f=1/2$, $T=T_c=4.5, L_z=2
L$.  Fig.\ \ref{rhofig} shows finite size scaling data collapses
according to Eq.\ (\ref{fss_rho}) of MC data for $\rho_x, \rho_z$ vs.\
the total MC integration time $t_0$ in the Kubo formula.  The inset
shows the RMS error in a power law fit to the MC data curves, and the
best fit is obtained for $z = 1.5 \pm 0.2$.  This gives a resistivity
exponent in $\rho \sim t^s$ of $s=\nu(z-1)\approx 0.3$.

\begin{figure}[htb]
\centerline{\epsfxsize 7cm\epsfbox{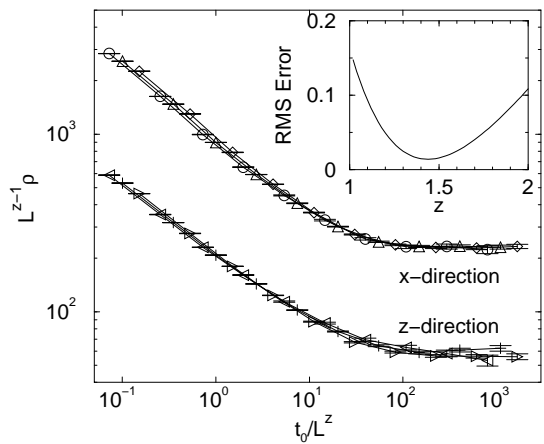}} \caption{Finite size
scaling data collapse of MC data for the resistivity, obtained for
$T_c=4.5$, $z=1.45$, and $L=8,10,12$. Inset: RMS fit error in the
data collapse.}
\label{rhofig}
\end{figure}

We will now discuss the values obtained for the critical exponents of
the random pinning model: $\zeta = 1 \pm 0.1, \nu = 0.7 \pm 0.1, z =
1.5 \pm 0.2$.  Our correlation length exponent $\nu$ is close to the
limiting value at a disordered fixed point allowed by the inequality
\cite{Harris}: $\nu \ge 2/d$.  Within the precision obtained, we can not
distinguish this exponent from that of the pure zero field 3D XY
model, i.e.\ $\nu \approx 0.67$.  Also our dynamic exponent $z \approx
1.5$ is consistent with the zero field 3D XY model with MC dynamics
for vortex loops \cite{Weber,dynamics}.  If confirmed, these results
suggest a scenario where the vortex glass transition in the random
pinning model for high fillings and strong disorder belongs to the
zero field 3D XY universality class.  In particular, this also
suggests that the glass transition is driven by similar effectively
isotropic, closed vortex loop fluctuations as in the zero field case,
on top of a glassy groundstate.

Finally we compare our results with other models for the vortex glass
transition and with experiments.  The critical exponents obtained here
for the random pinning model differ considerably from those of the
gauge glass model \cite{Olson-Young}: $\nu\approx 1.39, z\approx 4.2,
s=\nu(z-1)\approx 4.5$, and also from a random coupling 3D XY model
with an applied field and open boundary conditions \cite{Kawamura}:
$\nu\approx 2.2, z \approx 3.3, s \approx 5.3$.  Hence, these models
appear to belong to different universality classes than the random
pinning model.  Experiments on (K,Ba)BiO$_3$ give $s\approx 3.9$, and
experiments on untwinned proton irradiated YBa$_2$Cu$_3$O$_{7-\delta}$
give $s\approx 5.3$ \cite{vortexglass-experiment}.  These experiments
show consistency with expected vortex glass behavior for tilting the
magnetic field away from the $c$-axis.  Unexpectedly, the exponents
disagree considerably with our results.  Further work is needed in
order to clarify the reasons for this discrepancy.

In summary, we have observed a finite temperature continuous vortex
glass transition by simulations and finite size scaling analysis of a
three-dimensional random pinning model.  The critical exponents,
$\zeta \approx 1, \nu \approx 0.7, z \approx 1.5$, are surprisingly
close to those of the zero-field pure 3D XY model, but disagree with
the gauge glass model and with some experiments.  These results
motivate further theoretical and experimental work.

We acknowledge very useful discussions with Pierre Le Doussal, Peter
Olsson, Stephen Teitel, and Peter Young.  This work was supported by
the Swedish Natural Science Research Council, STINT, and PDC.


\begin{references}

\bibitem{Fisher} M.\ P.\ A.\ Fisher, \prl {\bf 62}, 1415 (1989); D.\
S.\ Fisher, M.\ P.\ A.\ Fisher, and D.\ A.\ Huse, \prb {\bf 43}, 130
(1991).

\bibitem{Olsson} P.\ Olsson and S.\ Teitel, \prl {\bf 87}, 137001
(2001).  Y.\ Nonomura and X.\ Hu, \prl {\bf 86}, 5140 (2001).

\bibitem{Bragg-glass} T.\ Giamarchi and P.\ Le Doussal, \prl {\bf 72},
1530 (1994); \prb {\bf 55}, 6577 (1997).

\bibitem{Olson-Young} T.\ Olson and A.\ P.\ Young, \prb {\bf 61},
12467 (2000).

\bibitem{vortexglass-experiment} T.\ Klein, A.\ Conde-Gallardo, J.\
Marcus, C.\ Escribe-Filippini, P.\ Samuely, P.\ Szab{\'o}, and A.\ G.\
M.\ Jansen, \prb {\bf 58}, 12411 (1998).  A.\ M.\ Petrean, L.\ M.\
Paulius, W.-K.\ Kwok, J.\ A.\ Fendrich, and G.\ W.\ Crabtree, \prl
{\bf 84}, 5852 (2000).

\bibitem{Kawamura} H.\ Kawamura, J.\ Phys.\ Soc.\ Jpn.\ {\bf 69}, 29
(2000).

\bibitem{screening} H.\ S.\ Bokil and A.\ P.\ Young, \prl {\bf 74},
3021 (1995).  J.\ Kisker and H.\ Rieger, \prb {\bf 58}, R8873 (1998).

\bibitem{Zimanyi} C.\ Reichhardt, A.\ van Otterlo, and G.\ T.\
Zimanyi, \prl {\bf 84}, 1994 (2000).

\bibitem{dynamics} J.\ Lidmar, M.\ Wallin, C.\ Wengel, S.\ M.\ Girvin,
and A.\ P.\ Young, \prb {\bf 58}, 2827 (1998).

\bibitem{helicity} J.\ Lidmar and M.\ Wallin, \prb {\bf 59}, 8451
(1999).

\bibitem{Kubo} A.\ P.\ Young, Proceedings of the Ray Orbach
Inauguration Symposium (World Scientific, Singapore, 1994).

\bibitem{Boseglass} J.\ Lidmar and M.\ Wallin, Europhys.\ Lett.\ {\bf
47}, 494 (1999).

\bibitem{dirtybosons} M.\ Wallin, E.\ S.\ Sorensen, S.\ M.\ Girvin,
and A.\ P.\ Young, \prb {\bf 49}, 12115 (1994).

\bibitem{exchange} K.\ Hukushima and K.\ Nemoto, J.\ Phys.\ Soc.\
Jpn.\ {\bf 65}, 1604 (1996).

\bibitem{Harris} A.\ B.\ Harris, J.\ Phys.\ C {\bf 7}, 1671 (1974).
J.\ T.\ Chayes, L.\ Chayes, D.\ S.\ Fisher, and T.\ Spencer, \prl {\bf
57}, 2999 (1986); Commun.\ Math.\ Phys.\ {\bf 120}, 501 (1989).

\bibitem{Weber} H.\ Weber and H.\ J.\ Jensen, \prl {\bf 78}, 2620
(1997).

\end{references}
\end{document}